\documentclass[twocolumn,prb,showpacs,preprintnumbers,amsmath,amssymb]{revtex4}
\usepackage[dvipdfm]{graphicx}
\usepackage{dcolumn}
\usepackage{bm}
\usepackage{amssymb}

\begin{document}
\title{Numerical study of magnetic field induced ordering in BaCuSi$_2$O$_6$
and related systems}

\author{Kwai-Kong Ng}
\author{Ting-Kuo Lee}
\affiliation{Institute of Physics, Academia Sinica, NanKang,
Taipei 11529, Taiwan}
\date{\today}
\begin{abstract}
Thermodynamics of spin dimer system BaCuSi$_2$O$_6$ is studied
using a quantum Monte Carlo calculation (QMC) and a bond-operator
mean field theory. We propose that a new type of boson, which,
rather than being hard-core, allows up to two occupancy at each
site, is responsible for the Bose Einstein condensation of field
induced ordering. Its superfluid density is identified as the
square of the in-plane staggered magnetization $m_{xy}$  in the
ordered phase. We also compare our QMC result of the spin
Heisenberg model to those predicted by mean field theory as well
as by the simple hard core boson model for both large and small
intra-dimer coupling $J$. The asymmetry of the phase diagram of
$m_{xy}(h)$ of small coupling $J$ in related systems such as
NiCl$_2$-4SC(NH$_2$)$_2$ is explained with our new boson operator.
\end{abstract}


\maketitle
\section{Introduction}
Recent experiment on BaCuSi$_2$O$_6$ under strong external
magnetic field observed a $\lambda$-like transition of the
specific heat capacity \cite{Jaime}. This is among other spin
dimer systems, such as KCuCl$_3$ and TlCuCl$_3$, that exhibit
quantum phase transition from spin liquid to a magnetically
ordered state with increasing magnetic field. The ordered phase is
characterized by an uniform magnetization accompanied by a long
range staggered magnetic order perpendicular to the field. The
staggered magnetization vanishes as temperature rises above the
transition temperature $T_c$. This has been interpreted as a
phenomenon of Bose-Einstein Condensation (BEC) of the magnetic
field induced $S_z = +1$ triplets, which, when ignoring higher
energy interactions, behave like repulsive hard-core bosons
\cite{Giamarchi}. The external field now plays the role of
chemical potential and controls the number of triplets.
Measurements of $T_c$ near critical field $h_c$ are reported to
have the deduced critical exponent $\nu=0.63(3)$ \cite{Sebastian},
in good agreement with the predicted value $\nu=\frac{2}{3}$ of
Bose-Einstein condensation. Numerical calculations also found the
critical exponent approaching the expected as $h \rightarrow h_c$
\cite{Sebastian,Nohadani} for 3D spin dimer systems, support the
notion of BEC.

Previous numerical calculation on the hard-core boson model found
a similar transition as observed \cite{Jaime}. Instead, in this
paper, we employ the original Heisenberg model to study the
temperature and field dependence of thermodynamic quantities. In
contrast to the hard-core model, we introduce a new type of boson,
which allows up to two occupancy at each site, and show that its
condensate density is identical to the staggered magnetization
$m_{xy}^2$ in the ordered phase with the global phase corresponds
to the direction of $m_{xy}$. The temperature and magnetic
dependance of $m_{xy}^2$ is explained in the context of
Bose-Einstein condensation. We also extend to the case with
smaller inter-plane coupling in which higher energy states are no
longer negligible in the ordered phase and leads to the asymmetry
of on both $m_z(h)$ and $m_{xy}^2(h)$. This condition is relevant
to other dimers material like TlCuCl$_3$ and $S=1$ system
NiCl$_2$-4SC(NH$_2$)$_2$ \cite{Paduan,Zapf}.

\section{Model Hamiltonian}
In BaCuSi$_2$O$_6$, while Cu$^{2+}$ ions arrange itself in layers
of square lattice, every two Cu-Si-O layers are separated by
planes of Ba$^{2+}$ ions, and the distance between adjacent
bilayers, 7.043 {\AA} \cite{Finger} , is much larger than the
bilayer separation, 2.73 {\AA}. Consequently, the spins of
Cu$^{2+}$ ions form dimers between bilayers and interact weakly in
plane and out of plane. We study the antiferromagnetic Heisenberg
model in an external field $h \hat{z}$ written as:


\begin{eqnarray}
H_{sp}&=&J \sum_i {\bf S}_{1,i} \cdot {\bf S}_{2,i} +
J_1 \sum_{\alpha,\langle i,j \rangle} {\bf S}_{\alpha,i} \cdot
{\bf S}_{\alpha,j} \nonumber \\
& &+ J_2 \sum_i {\bf S}_{1,i} \cdot {\bf S}_{2,i+\hat{z}} -\mu_B
g_0 h \sum_{\alpha,i} S^z_{\alpha,i},
\end{eqnarray}
where $\alpha=1,2$ denotes the layer index and the second
summation refer to summing over all nearest neighbors in the $xy$
plane for both types of layers. The exchange coupling constants
are taken as $J=4.45$ meV, $J_1$=0.58 meV and $J_2$=0.116 meV,
which is provided by ref. \onlinecite{Jaime} whose QMC based on
the hard-core boson model using these parameters yield the best
fit to the experimental result. The magnetic field $h$ yields the
Zeeman energy in the Hamiltonian and the gyromagnetic constant
$g_0=2.306$ in the case of BaCuSi$_2$O$_6$. We first focus on the
condition that $J\gg J_1,J_2$ so that the zero field ground state
only composed of inter-plane singlet dimer states. Finite energy
is needed to excite a singlet state
$|s\rangle\equiv({|\uparrow\downarrow\rangle}-|\downarrow\uparrow\rangle)/\sqrt{2}$
to a triplet state $|t_{+}\rangle \equiv |\uparrow \uparrow
\rangle$. Increasing the magnetic field $h$ will reduce the energy
gap which closes up at a critical $h_c\sim 24$ T such that $\mu_B
g_0 h_c$ is of order $J$. For $h > h_c$, a new magnetic order with
staggered magnetization in the $xy$ plane emerges, with the ground
state a linear combination of $|s\rangle$ and $|t_{+}\rangle$ in
each dimer. The breaking of rotational symmetry in the $xy$ plane
implies the existence of a Goldstone mode at {\bf
Q}=($\pi$,$\pi$,$\pi$). Raising the field $h$ further leads to a
finite transition temperature $T_c$ which reaches a maximum before
falls down again to zero at a saturation field $h_s$.

From another point of view, the quantum phase transition at $h_c$
can be considered as BEC of bosons $b^\dag \equiv t^\dag_{+1}$ on
the vacuum composed of $|s\rangle$. Since no more than one triplet
is allowed in the same dimer, the triplet dimers $b^\dag$, with
{\bf S}=+1, behave as hard-core bosons. These bosons interact
repulsively because $J_1$ and $J_2$ favor antiferromagnetic
couplings of neighboring dimers and hence neighboring bosons are
avoided. One can reduce $H_{sp}$ to an effective Hamiltonian
$H_{b}$ \cite{Jaime} of hard core bosons by projecting out higher
energy states $|t_0\rangle \equiv (|\uparrow \downarrow \rangle +
|\downarrow \uparrow \rangle) /\sqrt{2} $ and $|t_{-}\rangle
\equiv |\downarrow \downarrow\rangle$,
%
%
\begin{eqnarray}
H_b&=&t_1 \sum_{\langle i,j \rangle} (b^{\dag}_i b_j + h.c.)
+ V_1 \sum_{\langle i,j \rangle} n_i n_j \nonumber \\
& &+ t_2 \sum_i
(b^{\dag}_i b_{i+\hat{z}} + h.c.) +V_2 \sum_i n_i n_{i+\hat{z}} \nonumber \\
& &+ \mu \sum_i n_i,
\end{eqnarray}
where $t_1=V_1=J_1/2$, $t_2=V_2=J_2/4$ and $n_i=b^\dag_i b_i$. The
chemical potential, $\mu=J-\mu_B g_0 h$, now depends on the
magnetic field $h$ and only when $h>h_c$, a finite concentration
of bosons arises and condensate forms in the ground state. Note
that this Hamiltonian preserves a particle-hole symmetry that the
phase diagram of $T_c(h)$ (or $n_0(h)$, the superfluid density),
must be symmetric about a $h_m$ where $T_c$ is a maximum at
$n_0=1/2$. Numerical calculations based on this hard-core boson
model reproduce thermodynamical quantities that agree well with
the experimental results and support this boson picture
\cite{Jaime}. Matsumoto $et$ $al.$ pointed out that, however, when
constructing a consistent mean field theory
\cite{Matsumoto,Sommer}, the high energy state $|t_{-}\rangle$
should be included. This is reflecting the process of annihilating
$|t_{+}\rangle$ and $|t_{-}\rangle$ at neighboring dimers while
creating a $|s\rangle$ and vise versa. This process becomes
important when the inter-dimer coupling $J_1$ and $J_2$ are
comparable to intra-dimer coupling $J$ and when $h$ is not much
larger than $h_c$. Therefore, to provide a consistent and general
description for all $J$ and $h$, we introduce the appropriate
boson operator $\tilde{b}^\dag$:
\begin{equation}
\tilde{b}^\dag_i = \frac{(-1)^i}{\sqrt{2}} \left( S^+_{1,i} -
S^+_{2,i} \right), \label{boson}
\end{equation}
with its operations on the spin states

\parbox{3.5cm}{
\begin{eqnarray*}
\tilde{b}^\dag_i |s\rangle_i &=& |t_{+}\rangle_i \\
\tilde{b}^\dag_i |t_{+}\rangle_i &=& 0 \\
\tilde{b}^\dag_i | t_{-}\rangle_i &=& |s\rangle_i
\end{eqnarray*}}
\parbox{3.5cm}{
\begin{eqnarray*}
\tilde{b}_i |t_{+}\rangle_i &=& |s\rangle_i  \\
\tilde{b}_i |s\rangle_i &=& |t_{-}\rangle_i  \\
\tilde{b}_i |t_{-}\rangle_i &=& 0
\end{eqnarray*}}

\noindent for even $i$ and an extra negative sign for odd $i$.
Here $|t_{+}\rangle_i $ is redefined as $-|\uparrow \uparrow
\rangle_i$, and the first and second arrows denote the spins of
layer 1 and layer 2 respectively. These yield the expectation
values $\langle s| \tilde{b}^\dag \tilde{b} | s\rangle = \langle
t_{+} | \tilde{b}^\dag \tilde{b} | t_+ \rangle = 1$ and $\langle
t_- |\tilde{b}^\dag \tilde{b}| t_- \rangle = 0$. Unlike $b^\dag$,
$\tilde{b}^\dag$ operates on a Hilbert space of vacuum $|0\rangle
\equiv |t_-\rangle$, while $| s\rangle$ and $|t_+ \rangle$
correspond to single and double occupied states of $\tilde{b}$
respectively. $|t_0\rangle$ is again decoupled from the other
states and not included in the Hilbert space.
 Although occupation greater than two at
each lattice site is again prohibited, $\tilde{b}$ boson is no
longer the simple hard core boson described by $b$. Note that the
vacuum is higher in energy and the lowest energy state $|s\rangle$
is fully occupied when $h < h_c$.

The modified effective Hamiltonian is now written as:
\begin{eqnarray}
H_{\tilde{b}}&=& \sum_i \epsilon ( \tilde{b}^\dag_i
\tilde{b}_i + \tilde{b}_i \tilde{b}^\dag_i ) + \mu \sum_i \tilde{n}_i  \nonumber\\
& & + t_1 \sum_{\langle i,j \rangle} (\tilde{b}^{\dag}_i
\tilde{b}_j + h.c.) + V_1 \sum_{\langle i,j \rangle} \tilde{n}_i \tilde{n}_j \nonumber \\
& & + t_2 \sum_i (\tilde{b}^{\dag}_i \tilde{b}_{i+\hat{z}} + h.c.)
+ V_2 \sum_i \tilde{n}_i \tilde{n}_{i+\hat{z}},
\end{eqnarray}
where $\epsilon=-J$, $\mu=-\mu_B g_0 h$ and
$\tilde{n}_i=\tilde{b}^\dag_i \tilde{b}_i - \tilde{b}_i
\tilde{b}^\dag_i$ with $t_{1,2}$ and $V_{1,2}$ unchanged. The
potential energy terms arise from the term $S^z_i S^z_j$ of the
original spin Hamiltonian $H_{sp}$ which is attractive for $t_+$
and $t_-$ but is repulsive for the same kind of triplets. This
Hamiltonian now loses the symmetry between $|s\rangle$ and
$|t_+\rangle$ due to the presence of $|t_-\rangle$. For $h_c < h <
h_s$, it is the condensation of this boson $\tilde{b}_i$ that
gives the quantum phase transition observed.

\section{Mean field theory}
A mean field condensate ground state of the effective Hamiltonian
$H_{\tilde{b}}$ is written as \cite{Matsumoto}
\begin{eqnarray}
|\Psi_0\rangle &=& \prod_{i} \left( u \tilde{b}_i^\dag + v
(-1)^i(f e^{i \theta} \tilde{b}_i^{\dag 2} +
g e^{-i \theta} ) \right) |0\rangle_i,  \nonumber \\
&=& \prod_{i} \left( u |s\rangle_i + v (-1)^i(f e^{i
\theta}|t_+\rangle_i + g e^{-i\theta}|t_-\rangle_i) \right)
\label{wf}
\end{eqnarray}
where $u^2+v^2=1$ and $f^2+g^2=1$ and all parameters are real.
The last equation is exactly the
 same as the one taken for mean field condensate using bond operator
representation \cite{Matsumoto} in which $\theta$ is chosen to be
zero. The global phase $\theta$ corresponds to the angle of
rotation in the $xy$-plane whose presence, due to the rotational
invariance, should not change the energy of the system.
 $t_+ $ and $t_-$
undergo  the transformation $t_+ \rightarrow e^{i\theta} t_+ $ and
$t_- \rightarrow e^{-i\theta} t_-$ when the $x$ and $y$ axes are
rotated by an angle $\theta$. As shown below, this phase $\theta$
also specifies the orientation of the in-plane staggered
magnetization.

Remarkably, the ground state $|\Psi_0\rangle$ has two finite
expectation values that correspond to two order parameters:
\begin{eqnarray}
\langle \tilde{b} \rangle &=& uv(f+g)e^{i\theta} \equiv
\tilde{b}_0 e^{i\theta} \label{ave_b}\\
\langle \tilde{b}^2 \rangle &=& v^2
fge^{2 i \theta}.
\end{eqnarray}
The first one is the usual order parameter expected from a BEC, while
the second additional one is originated from the fact that $\tilde{b} $
allows up to two bosons occupied at each site $i$. These expectation values
can be related to the staggered magnetization and spin-spin correlation
function when Eq. \ref{boson} is applied.

In the spin language, the staggered magnetization is taken as the
order parameter for the ordered phase and defined as
\begin{eqnarray*}
m^2_{xy} &=& m^2_x + m^2_y,  \\
m_\alpha &=& \frac{1}{N\sqrt{2}} \sum_i (-1)^i \langle
S^\alpha_{1,i} - S^\alpha_{2,i} \rangle \quad \alpha=x,y.
\label{mxy}
\end{eqnarray*}
As mentioned above, one can easily show that $m_{xy}$ is identical
to the magnitude of superfluid order parameter $\tilde{b}_0$.
Using the definition of $\tilde{b}^{\dag}$ from Eq. \ref{boson},
one obtains
\begin{eqnarray*}
m_x &=& \frac{1}{2N } \sum_i \langle \tilde{b}^\dag_i +
\tilde{b}_i \rangle \\
&=& \tilde{b}_0 \cos{\theta}.
\end{eqnarray*}
Similarly,
$m_y = \tilde{b}_0 \sin{\theta}$ and
consequently,
\begin{equation}
m_{xy}= \tilde{b}_0=uv(f+g),
\end{equation}
and the phase $\theta$ specifies
the direction of $m_{xy}$. The realization of $m_{xy}= \tilde{b}_0$
allows us to compute the staggered magnetization directly using QMC.

The uniform magnetization along $\hat{z}$ axis $m_z$ can also be written as
\begin{eqnarray}
m_z &=& \frac{1}{N} \sum_i \langle \tilde{b}_i^\dag \tilde{b}_i -
\tilde{b}_i \tilde{b}_i^\dag \rangle \nonumber\\
& =& v^2 (f^2-g^2),
\end{eqnarray}
which is simply the difference between the number of $|t_+
\rangle$ and of $|t_- \rangle $.

We stress that
the distinction between $\tilde{b}^\dag$ and $ b^\dag$ originates from
the consideration of $|t_- \rangle$, which we take as the vacuum here.
 If the $|t_- \rangle $ state is much higher in energy and being ignored,
$\tilde{b}^\dag$, as $g \rightarrow 0$, is just the hard core
boson $b^\dag$ described in ref. \onlinecite{Jaime}. This
approximation becomes exact when $J \rightarrow \infty$, but will
be insufficient when $J \sim J_1, J_2$.

A complete description of the static and dynamical properties  of
BaCuSi$_2$O$_6$ under applied field can be obtained by performing
a mean field analysis using the bond operator representations. The
same method has been successfully applied to related compounds of
KCuCl$_3$ and TlCuCl$_3$ by Matsumoto $et$ $al. \cite{Matsumoto}$

Under bond operator representations, each spin operator is
replaced by boson operators, one singlet and three triplet
operators, that operate on a inter-plane dimer bond and the
original spin Hamiltonian is then transformed to a Hamiltonian of
interacting bosons.   These are hard core bosons because of the
constraint $1=s^\dag s+t_+^\dag t_+ + t_0^\dag t_0 + t_-^\dag
t_-$. Two unitary transformations are performed to mix the bond
operators in the ordered phase to give the appropriate condensate
$\bar{a}=\langle a_i \rangle_{av}$,
\begin{equation*}
a_i=u s_i + v e^{-i \bf{Q}\cdot\bf{r}_i} (f e^{i\theta} t_{+i} + g
e^{-i \theta} t_{-i}).
\end{equation*}
For convenience we set $\theta=0$. This ground state condensate is
in fact identical to $|\Psi_0\rangle$. The mean field approach
proceeds as usual by taking the $a_i$ as a uniform field $\bar{a}$
and minimizing the energy to obtain a set of self-consistent
equations. The parameters $u$, $v$, $f$ and $g$ are now determined
by the self-consistent equations. The particle number constraint
does pose a problem on the decoupling of operators in the above
procedure. Naively employing a Lagrange multiplier to account for
the constraint would not close the energy gap in the ordered
phase. In ref. \onlinecite{Matsumoto} a Holstein-Primakoff
approximation is taken instead by assuming the contribution of
energy modes other than $a_i$ is small in the ground state. This
approach successfully recovered the Goldstone mode with expected
features for the ground state. We refer the readers to ref.
\onlinecite{Matsumoto} for the details and only present our result
here.

\section{Quantum Monte Carlo calculation}
The aim of quantum Monte Carlo simulations is twofold: to test the
validity of the proposed mean field theory and to justify the
hard-core boson model. We stress that the simulations are
performed on the original spin Heisenberg model rather than the
boson model and any observed phase transition is a pure
consequence of spin exchange coupling of $H_{sp}$.

Having been successfully applied to spin systems in external
magnetic field in the last decade and demonstrated its advantages
over the standard worldline approach, the stochastic series
expansion (SSE) \cite{Sandvik} is the method of choice for our
problem. The algorithm of measuring Green's function introduced by
Dorneich $et$ $al.$ \cite{Dorneich} allows us to compute the spin
correlations like $\langle S^+_i S^-_j \rangle$. In this scheme,
however, since spin must be conserved in the ensemble, direct
measurement of staggered magnetization $m_{xy}$, which requires
the knowledge of $\langle S^+_i \rangle$ and $\langle S^-_i
\rangle$, is impossible. Instead, in our calculation, we compute
the superfluid density
\begin{equation}
n_0 = \frac{1}{N^2} \sum_{i,j} \langle \tilde{b}^\dagger_i
\tilde{b}_j \rangle.
\end{equation}
In MFT, Eq. \ref{ave_b} and Eq. \ref{mxy} lead to
$n_0=\tilde{b}^2=m_{xy}^2$, and to which our simulated $n_0$ can
be compared. Using the definition of $\tilde{b}$, Eq. \ref{boson},
the computation of $n_0$ involves only measuring spin correlations
which can be done easily with Dorneich's method. Therefore a
direct comparison of the calculated order parameter by QMC to
those predicted by mean field approximation as well as
experimental data is possible.

All simulations are performed on cubic lattice of size 12x12x12
typically with 4x10$^5$ update measurement cycles for each data
point, except the measurement of specific heat which requires up
1.6x10$^6$ update cycles to achieve the acceptable accuracy.

\section{Thermodynamics}

\begin{figure}
\includegraphics[width=\linewidth]{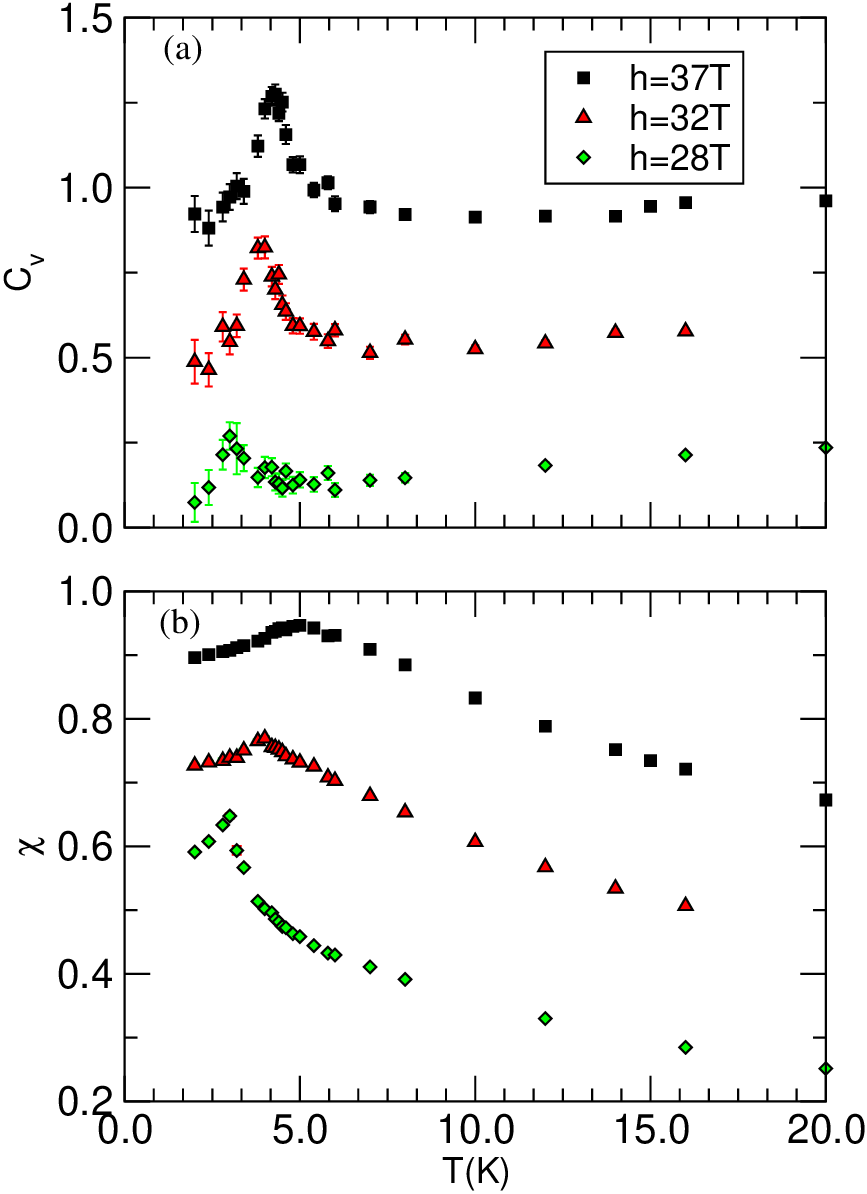}
\caption{(color online) QMC result of (a) specific heat $c_v$ and
(b) spin susceptibility $\chi$ vs. temperature for different
magnetic fields. Data for different $h$ are shifted for clarity.}
\label{fig1}
\end{figure}

The specific heat and susceptibility at finite temperature around
the transition have been experimentally measured by Jaime {\it et
al.} \cite{Jaime} for various magnetic fields. The sharp peak of
specific heat at $T_c$ has a familiar $\lambda$ shape similar to
the one found in liquid helium He$^4$. Our QMC calculations based
on the spin Hamiltonian $H_{sp}$ reproduce this result for
different $h$ as shown in Fig. \ref{fig1}(a). A peak of the
specific heat $c_v$ signals the BEC develops when $h > h_c$ ($h_c
\sim 24$ T) and the peak grows with increasing $T_c$ as $h$ is
increased. The calculated values of $T_c$ approximately agree with
the experimental result. For $h=37$ T, our calculated and the
experimental $T_c$ are $\sim 4$ K and $\sim 3.75$ K respectively
\cite{Jaime}. Remind that we adopted the parameters $J$, $J_1$ and
$J_2$ used in ref. \onlinecite{Jaime} in which the hard-core boson
Hamiltonian $H_b$ was used to reproduce the experimental result.
Therefore the discrepancy on $T_c$ between QMC and experiment is a
consequence of ignoring higher energy states in $H_b$. One can
certainly produce a better fit to the experimental data if
different couplings than those obtained by ref. \onlinecite{Jaime}
are used. 

The calculated spin susceptibility $\chi_s$ in Fig. \ref{fig1}(b)
also shows a peak at $T_c$. The formation of boson condensate
below $T_c$ freezes the spins in the condensate and therefore
reduces the response to the external field that leads to a drop in
$\chi_s$. Unlike the specific heat, the transition peak reduces
when $h$ increases and becomes rather flat at $h\sim h_m$. 
\begin{figure}
\includegraphics[width=\linewidth]{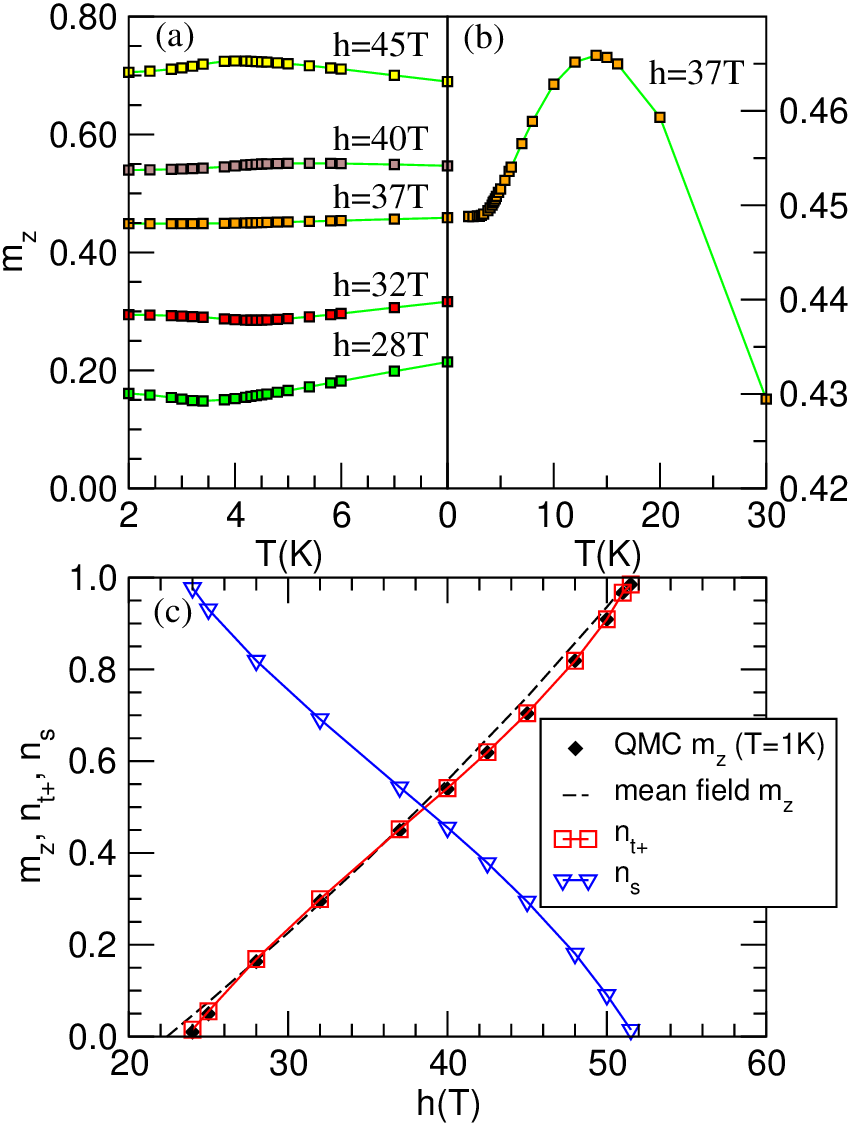}
\caption{(color online) (a) Low temperature regime of uniform
magnetization per dimer $m_z$ for different magnetic fields that
exhibits BEC. (b)
 The whole temperature range of $m_z$ for field $h=37$ T.
(c) Magnetic field dependence of $m_z$ obtained from both QMC
(T=1K) and MFT together with the number density of $s$ and $t_+$
states. } \label{fig2}
\end{figure}

The phase transition
is also visible in the low temperature regime of uniform magnetization
in which a minimum (maximum) is reached
 at $T_c$ for $h < h_m$ ($h > h_m$) as displayed in Fig. \ref{fig2}(a).
For $h < h_m$, the low-lying energy states have the $|s \rangle$
state dominating and so as $T$ approach $T_c$ from above, $m_z$,
which is approximately equal to the number of $t_+$ states,
reduces. However, below $T_c$, due to the formation of superfluid
condensate, number of $t_+$ states increases as the fraction of
condensate grows and therefore raises the $m_z$. This feature is
also found in the attractive Hubbard model of electrons
\cite{Singer} in which the double occupancy, i.e. the number of
on-site Cooper pairs, increases as T approaches 0 from $T_c$.

The role of $s$ and $t_+$ states exchanges for $h>h_m$ and a
maximum of $m_z$ is resulted instead. The same temperature
dependence of magnetization was also observed in another dimer
system TlCuCl$_3$ \cite{Oosawa} experimentally and a Hartee-Fork
calculation \cite{Nikuni} on hard core dilute magnons explained
the qualitative feature of $m_z(T)$ around $T_c$ of the proposed
BEC. QMC simulations on different coupling strength \cite{Wessel}
has shown that this is a general feature of 3D coupled dimer
system while the extrema are missing in non-interacting dimers.

Since the total uniform magnetization is simply the difference
between the number of $|t_+\rangle$ state and of $|t_-\rangle$,
further increase of temperature (to $T\sim 15$ K for $h=37$ T)
will raise the number of $|t_-\rangle$, as well as $|t_0\rangle$
and finally suppresses the magnetization $m_z$ as shown in Fig.
\ref{fig2}(b).

\begin{figure}
\includegraphics[width=\linewidth]{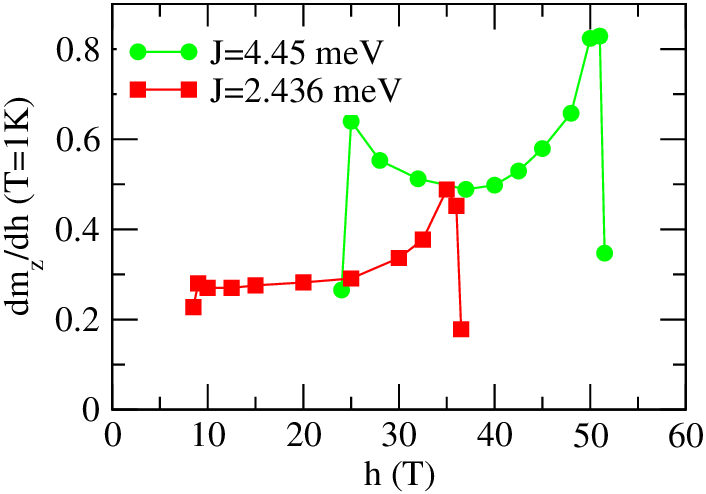}
\caption{(color online) The slope of $m_z(h)$, i.e. the
susceptibility, at $T=1$K.} \label{fig3}
\end{figure}

In Fig. \ref{fig2}(c) we plot the field dependence of $m_z$
obtained by both QMC and MFT, which increases monotonically as $h$
increases, consistent to experimental results. While $m_z(h)$ is
more or less linear in the mean field case, its slope reduces and
then increases in the QMC calculation. This change of slope is
again a consequence of the formation of condensate. Without the
condensation, $m_z(h)$ will be simply a straight line at $T=0$.
But the formation of condensate below $T_c$ increases $m_z$ for $h
< h_m$ as discussed above. While the minimum of $m_z(T)$
disappears at $h_c$ ($T_c=0$) and $h_m$, there is no increase of
magnetization due the condensation at these two fields. Therefore
the resulted $m_z(h)$ curve for $h<h_m$ must be convex as shown.
Similar argument applies to the high field regime $h>h_m$ and a
concave curve is predicted. The slope of $m_z(h)$, that is the
susceptibility, is shown in Fig. \ref{fig3}. The small asymmetry
indicates the existence of $t_-$ states which becomes more
significant for smaller intra-dimer coupling $J$ as will be
discussed in the next section.

 Also plot in Fig. \ref{fig2}(c) is
$n_{t+}$ and $n_s$, the average number of triplet $t_+$ and
singlet $s$ per dimer respectively. In the MFT, $n_{t+}=v^2 f^2$,
$n_{t-}=v^2 g^2$ and $n_s=u^2$, where $u$, $v$, $f$ and $g$ are
parameters given by the condensate ground state $|\Psi_0\rangle$
(Eq. \ref{wf}). $n_{t-}$ is found to be negligible and therefore
is not shown in Fig. \ref{fig2}(c). While $n_s$ reduces as
$n_{t+}$ increases, their total number is almost 1 (less than 2
percent in difference) for all fields. The intercept of both
curves gives an estimate of $h_m=38.6$ T in which $n_{t+} \approx
n_s \approx 0.5$.

The fact that $n_{t+}$ essentially coincides with $m_z$ means the
average number of $t_-$ is almost zero ($g^2$ is very small) in
BaCuSi$_2$O$_6$. This explains why the hard-core boson model $b$
gives results consistent to experimental data. However, the
derivation from the hard-core boson model will be more apparent
when we discuss the $m_{xy}$ where the difference is proportional
to $g$ instead of $g^2$. 
\begin{figure}
\includegraphics[width=\linewidth]{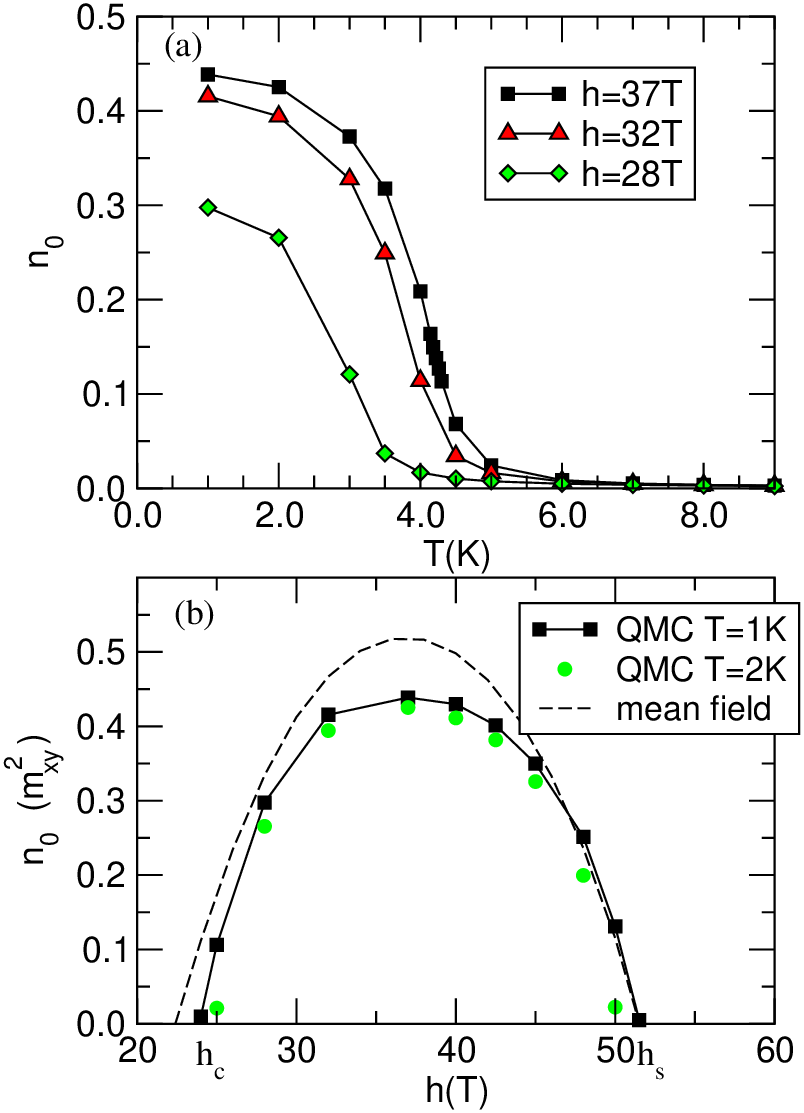}
\caption{(color online) Superfluid density $n_0$, which is
equivalent to $m_{xy}^2$ in the MFT, as (a) a function of
temperature and (b) a function of magnetic fields. } \label{fig4}
\end{figure}

Temperature dependence of $n_0$ is shown in Fig. \ref{fig4}.
$n_0$, equivalent to the superfluid density $m_{xy}^2$ in MF,
 starts to rise at about the same $T_c$ obtained from
specific heat and spin susceptibility and therefore justifies the
idea of BEC of $\tilde{b}^\dag$ for the transition observed in
BaCuSi$_2$O$_6$. The broken rotational symmetry below $T_c$ that
generates $m_{xy}$ is identical to the broken gauge invariance of
$|\Psi_0\rangle$. Also shown in Fig. \ref{fig4} is the field
dependence of $m_{xy}^2$ at low temperature. The data shows that
although MF calculation overestimates the order parameter near the
field $h_m$, in general it agrees well with QMC result. The
predicted MF values of critical and saturation fields are given
by:
\begin{eqnarray*}
\mu_B g_0 h_c &=& \sqrt{J(J-4 J_1-2J_2)} \\
\mu_B g_0 h_s &=& J+4 J_1 + 2J_2.
\end{eqnarray*}
Since quantum fluctuation is totally surpassed in the fully
polarized phase, mean field theory predicts the exact saturated
field $h_s$ as expected. Below $h_s$, quantum fluctuation
stabilizes the disordered state and so reduces the order parameter
away from the MF value. The small discrepancy around $h_c$ will
become more significant when we lower the inter-plane coupling $J$
where MF gives a much smaller $h_c$ than the true value. It is
also obvious that $m_{xy}^2$ is asymmetric  about $h_m$, contrast
to what is predicted by the hard-core boson model that retains a
particle-hole symmetry. This indicates that $t_-$ states, although
the average number is small, does play a measurable role in the
ordered phase.

\section{Reduced exchange coupling $J$}
\begin{figure}
\includegraphics[width=\linewidth]{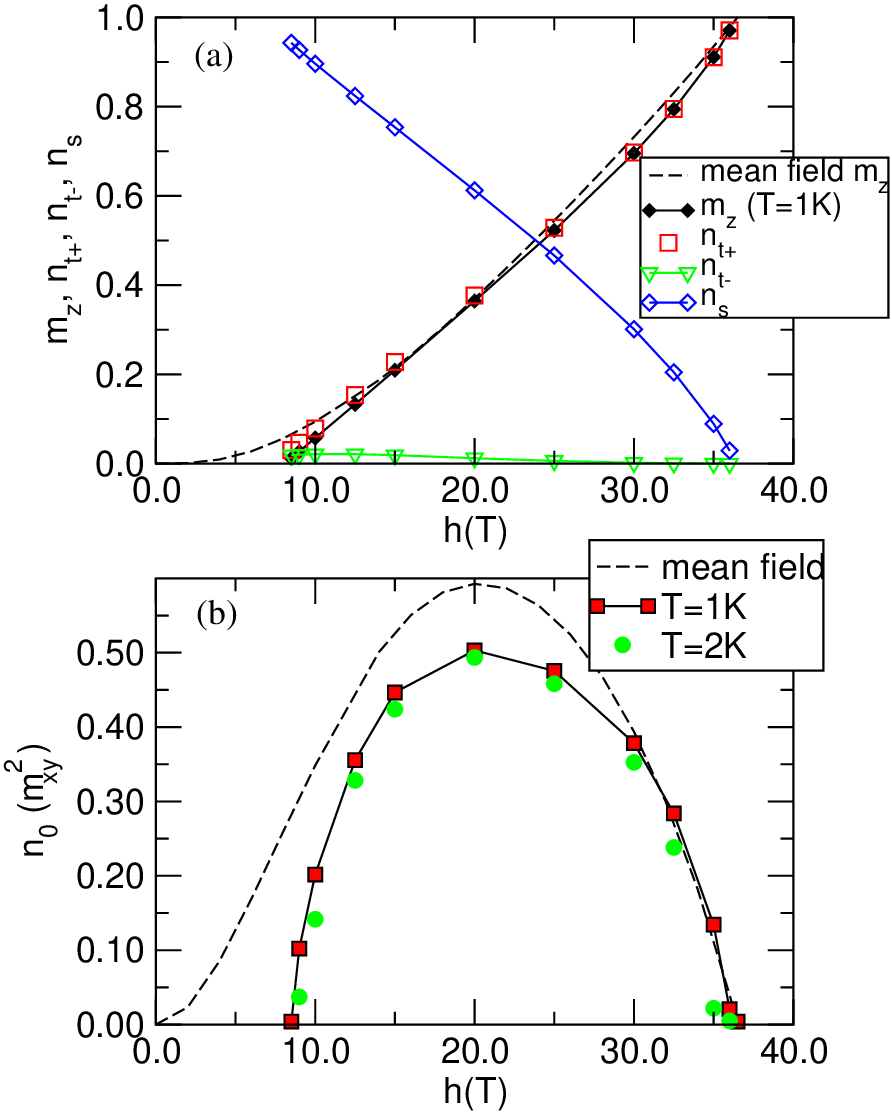}
\caption{(color online) Results of reduced inter-plane coupling
$J=2.436$ meV. (a) Temperature dependence of $m_z$ and number
density of $s$, $t_+$ and $t_-$ states. (b) Phase diagram of
$n_0(h)$.} \label{fig5}
\end{figure}

In BaCuSi$_2$O$_6$, since $J_1/J$ and $J_2/J$ are relatively
small, the mixing of $|t_-\rangle$ state is not significant, as
least for $h \gg h_c$. But there are other dimer systems, like
TlCuCl$_3$, where inter-dimer coupling is strong enough that
significant mixing of $|t_-\rangle$ is expected and new features
are possible. Therefore we extend our study to smaller intra-dimer
coupling $J$ on the same bilayer lattice as in BaCuSi$_2$O$_6$.
One possibility to reduce $J$ experimentally is the substitution
of Si by atoms of larger radius but the same chemical valence, in
order to enlarge the intra- dimer distance. Germanium, the one
right below Si in the period table, is the most natural candidate.

Reducing $J$ also implies the reduction of $h_c$ because triplet
states have lower energies with small $J$. We choose the reduced
$J$ to be $J=2.436$ meV such that the critical field $h_c$
predicted by MFT is exactly zero. The calculated curve (Fig.
\ref{fig5}) of $m_z(h)$ of MFT and QMC lay close to other each
although QMC yields a finite $h_c$ instead of zero. The average
number of $t_-$ is not negligible any more in this case. $n_0$, or
$m_{xy}$, shows a much larger discrepancy between QMC and MFT than
the large coupling $J$ case. MFT overestimates $m_{xy}$ for a
large range of magnetic field and   only when close to the
saturation field $h_s$ that MFT is close to exact results. The
asymmetry of $m_{xy}$ is also enhanced when $J$ is reduced as
expected. Although $n_-$ is still very small that $m_z$ deviates
only slightly from $n_+$, the square root of it, $\sqrt{n_-}=v g
$, can be as large as 0.14 and hence implies that $H_b$
 is not valid any more. The change of slope of $m_z(T)$ is also observable
 but is significantly asymmetric around $h_m\sim 20$ T as shown in Fig. \ref{fig3}.

Here we discuss how our results of reduced $J$ can be successfully
applied to a related compound NiCl$_2$-4SC(NH$_2$)$_2$
(DTN)\cite{Paduan,Zapf}. Although this material is not a spin
dimer system, the single ion anisotropy ($D\sim 0.88$ meV) plays
the role $J$ and splits the Ni $S=1$ spin state into $S_z=0$ and
$S_z=\pm1$ states, which corresponds to $s$ and $t_{\pm}$ states
in the above discussion. Under an external magnetic field the same
field induced ordering has recently been observed experimentally
in DTN\cite{Paduan,Zapf}. In this case, our proposed condensate
wave function becomes exact due to the absence of a corresponding
$t_0$ state. Therefore the field induced transition observed in
DTN shares the same physics as in spin dimer systems just
described. Since $D\sim 0.88$ meV is comparable to the average
exchange coupling $J'\sim 0.66$ meV, a large asymmetry in the
phase diagram is expected as we considered in the dimer system.
Recent experimental data indeed showed strong asymmetry in
$T_c(h)$ as well as in the change of slope of magnetization curve.
Instead of the proposed lattice expansion in the compound
\cite{Zapf}, the anomalous change of slope in magnetization is in
fact a consequence of the formation of superfluid condensate which
either enhances or reduces $m_z$ according to the applied field.

\section{Summary}
Working on the original antiferromagnetic Heisenberg model, our
QMC calculations reproduce the experimental finding of the field
induced ordering observed in both specific heat and spin
susceptibility. The calculated $c_v$ shows a $\lambda$-like
transition at $T_c$ with the amplitude maximizes at $h\sim h_m$
and reduces to zero as $h$ approaches $h_c$ or $h_s$. This
phenomenon can also be described as a Bose-Einstein condensation
of magnons. Instead of hard-core boson, we propose a
semi-hard-core boson, for which up to two occupancy is allowed, is
responsible for the field induced condensation.  We showed that
the superfluid density $n_0$ is identical to the in-plane
staggered magnetization $m_{xy}^2$, which is the order parameter
in the spin language. Our QMC results show a rise of
$n_0=m_{xy}^2$ at the same $T_c$ of $c_v$ and justify the idea of
Bose-Einstein condensation. Due to the smallness of $J_1$ and
$J_2$ in BaSiCu$_2$O$_6$, the number of $t_-$ states is tiny and
leads to the success of hard-core boson model $H_b$. Nevertheless,
the inadequacy of $H_b$ is still observable from the asymmetry of
the phase diagram. A bond operator mean field approach agrees well
with QMC data of $m_z$ for all field, and of $m_{xy}^2$ for
magnetic field close to $h_c$ and $h_s$. Quantum fluctuations that
ignored in the MFT lead to an overestimated $m_{xy}^2$ around
$h_m$ however. The difference is more significant as one considers
a smaller inter-plane coupling $J$. MFT predicts a much smaller
$h_c$ than the value found by the QMC. The asymmetry of $m_{xy}^2$
is also much enhanced and signals the failure of simple hard-core
boson model. This is indeed the case observed in
NiCl$_2$-4SC(NH$_2$)$_2$. The condensation of magnons also leads
to the change of slope in $m_z(h)$ which is again symmetric around
$h_m$ when $J$ is large but becomes asymmetric for smaller $J$
systems such as observed in NiCl$_2$-4SC(NH$_2$)$_2$.

In the ordered phase, the bond operator mean field theory
successfully account for longitudinal fluctuation, which is absent
in the conventional mean field theory, in addition to the
transverse fluctuation. The longitudinal fluctuation, or the spin
amplitude mode, mixes with the transverse fluctuation, the phase
mode, at lower field but becomes dispersionless and separated from
the phase mode at higher field. However, topological excitation,
corresponds to vortices of the condensate, can not be described in
the framework of bond operator MFT and further work will be
needed. Recently a BEC of magnons in frustrated triangular lattice
of antiferromagnet has been proposed and a similar
$\lambda$-transition in specific heat is also observed
\cite{Radu}. It raises the question whether the BEC is a more
general phenomenon of quantum antiferromagnet in external field
and also raises the possibility of some novel phenomena arising
from quantum nature of the condensate.

\begin{acknowledgments}
 We acknowledges financial support by the NSC
(R.O.C.), grant no. NSC 93-2112-M-001-018.
\end{acknowledgments}

\end{document}